\documentclass[aps,prb,reprint,superscriptaddress,10pt]{revtex4-2}
\usepackage{amsmath,makecell,multirow}
\usepackage{amssymb}
\usepackage{bm}
\usepackage{mathtools}
\usepackage{hyperref}
\usepackage[version=4]{mhchem}
\AtBeginDocument{\selectfont}
\usepackage[dvipsnames]{xcolor}

\hypersetup{
	hidelinks, colorlinks = true, linkcolor = blue, citecolor = Red, urlcolor = blue
}
\makeatletter
\renewcommand{\fnum@figure}{\textbf{\figurename~\thefigure}}
\renewcommand{\fnum@table}{\textbf{\tablename~\arabic{table}}}
\makeatother
\renewcommand{\figurename}{Fig.}
\renewcommand{\tablename}{Table}

\begin{document}
	
	\title{Hydration-Controlled Layer Stacking in \\\ce{(NH3)2Cu5(SeO3)2(OH)6\cdot(H2O)_{2+x}} ($x$ = 0, 1, and 3)}
	
	\author{Priya R. Baral}
	\affiliation{Crystal Growth Facility, Institute of Physics, École Polytechnique Fédérale de Lausanne, CH-1015 Lausanne, Switzerland}
	
	\author{Christian Jandl}
	\affiliation{ELDICO Scientific AG, c/o Switzerland Innovation Park Basel Area AG, Hegenheimermattweg 167 A, 4123, Allschwil, Switzerland}
	
	\author{Pauline Pradal}
	\affiliation{Crystal Growth Facility, Institute of Physics, École Polytechnique Fédérale de Lausanne, CH-1015 Lausanne, Switzerland}
	
	\author{Johann Roos}
	\affiliation{Crystal Growth Facility, Institute of Physics, École Polytechnique Fédérale de Lausanne, CH-1015 Lausanne, Switzerland}
	
	\author{Wenhua Bi}
	\affiliation{Crystal Growth Facility, Institute of Physics, École Polytechnique Fédérale de Lausanne, CH-1015 Lausanne, Switzerland}
	
	\author{Arnaud Magrez}
	\email{arnaud.magrez@epfl.ch}
	\affiliation{Crystal Growth Facility, Institute of Physics, École Polytechnique Fédérale de Lausanne, CH-1015 Lausanne, Switzerland}

	\date{\today}
	
	\begin{abstract}
		Hydration and dehydration are powerful yet underexplored variables for controlling the architecture of layered inorganic materials, because intercalated water can modify interlayer separation, hydrogen-bonding networks, and layer stacking. Here, we report the reflux synthesis of a new family of hydrated layered copper selenites, \ce{(NH3)2Cu5(SeO3)2(OH)6\cdot(H2O)_{2+x}} ($x$ = 0, 1, 3). From the crystal structures determined using electron diffraction and single crystal X-ray diffraction, we deduce that all three compounds share an identical layer built from Cu(OH)$_4$ squares and Cu-centered square pyramids forming distorted kagomé-like Cu$^{2+}$ network. While the intralayer atomic arrangement is preserved across the series, the degree of hydration governs both the interlayer separation and the stacking sequence. These compounds therefore provide a rare platform relevant to the design of hydration-responsive materials for sensing, ion transport, separations, actuation, and energy-related applications. The preservation of distorted kagomé-like Cu$^{2+}$ layers across hydration states further suggests potential interest for examining how interlayer water and stacking sequence affect low-dimensional magnetic coupling. Under reflux conditions, these phases are also shown to act as reactive intermediates in the formation of \ce{Cu2OSeO3}, establishing them as tunable precursors for copper oxoselenite synthesis.
		
	\end{abstract}
	
	\maketitle
	
	\noindent\textbf{Introduction.} Copper selenites and selenates constitute a structurally versatile class of inorganic oxosalts in which crystal chemistry is governed by the coordination flexibility of Cu$^{2+}$ and the bonding preferences of selenium oxoanions. Cu$^{2+}$ can adopt diverse coordination environments, while Se$^{4+}$ ions, with their stereochemically active lone pair, and Se$^{6+}$ ions, in tetrahedral selenate groups, promote an exceptional diversity of structural motifs, from isolated polyhedra, chains, layered networks, to three-dimensional frameworks~\cite{escamilla2003effect,effenberger1986kristallstrukturen,snyman1964polymorphism,effenberger2023cuseo4,becker2006reinvestigation,kovrugin2017synthesis,effenberger1986,arakcheeva2025cu2oseo3,kolitsch2001copper,effenberger1985cu,lafront1995layered,canossa2018structural,iskhakova1995crystal,keding1984crystal,giester1999new,giester1991crystal,gurzhiy2015cu3}. This structural richness makes copper selenium oxosalts attractive systems for exploring how local coordination, hydrogen bonding, and supramolecular organization determine the architecture of inorganic materials.
	
	In layered inorganic compounds, hydration provides a particularly powerful structural variable. Water molecules can occupy interlayer spaces, modify hydrogen-bonding networks, expand interlayer distance, and alter the relative stacking of adjacent layers. Conversely, dehydration can reduce interlayer separation, suppress water-mediated hydrogen bonds, and promote layer translations or rearrangements. Such hydration- and dehydration-driven transformations are well established in layered systems such as clays~\cite{suquet1987parameters}, layered double hydroxides~\cite{iyi2007factors}, graphene oxide~\cite{ahn2024two}, where variations in water content can produce large changes in structure and properties.
	
	Hydration-responsive layered materials are attractive because reversible water intercalation provides a direct means of coupling structure to function. Changes in interlayer distance, hydrogen-bond topology, dielectric response, mechanical strain, and ion mobility underpin application directions such as humidity sensing~\cite{an2019water}, selective ion transport and membrane separations~\cite{anwar2025recent}, moisture-driven actuation~\cite{zhang2024research}, ion exchange and pollutant capture and controlled molecular release~\cite{fang2022new}, and electrochemical or thermochemical energy storage~\cite{park2020water}. In this context, hydration-controlled stacking is not only a crystallographic phenomenon, but also a design principle for materials whose properties can be tuned by water activity.
	
	\begin{figure*}[htb!]
		\begin{center}
			\includegraphics[width=0.8\linewidth]{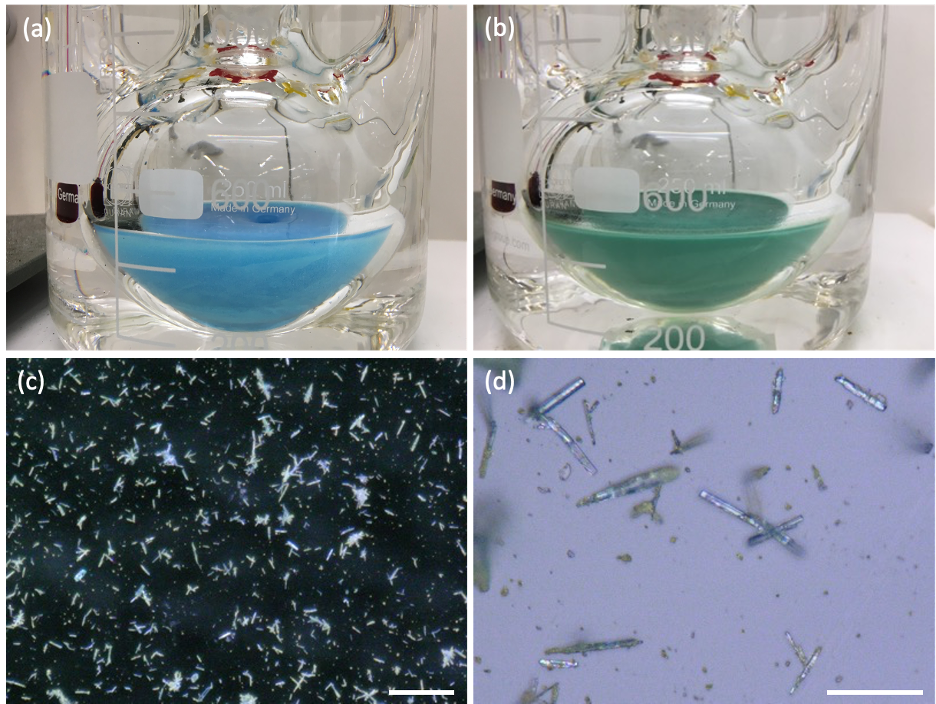}
			\caption{Reflux synthesis and single crystals of the intermediate phases. (a) Optical image of \ce{CuSeO3}$\cdot$2H$_2$O under reflux in ammonia containing solution. (b) After few minutes a color change of the solid is observed. (c) Low magnification (scale bar: 150~$\mu$m) and (d) high magnification (scale bar 50~$\mu$m) optical images of the \ce{(NH3)2Cu5(SeO3)2(OH)6\cdot(H2O)3} needles. Small green particles of \ce{Cu2OSeO3} are found beside as well as on the needles’ surfaces.}
			\label{fig1}
		\end{center}
	\end{figure*}
	
	Despite this broader relevance, the Cu–Se–N–O–H chemical space remains largely unexplored. Only five crystal structures composed exclusively of Cu, Se, N, O, and H are currently catalogued in crystallographic databases, and only one of these is a selenite derivative. In \ce{(NH4)3Cu(HSeO3)2(NO3)3}, square-planar CuO$_4$ units are linked by (HSeO$_3$)$^{2-}$ groups into layers separated by ammonium and nitrate ions~\cite{lafront1995layered}. The remaining compounds are selenate derivatives. In \ce{Cu(NH3)4SeO4}, Cu$^{2+}$ adopts a square-pyramidal coordination geometry with four \ce{NH3} ligands in the equatorial plane and an apical SeO$_{4}^{2-}$ group; the resulting chains are held together by hydrogen bonds between ammine and selenate groups of neighboring chains~\cite{morosin1969crystal}. \ce{(NH4)Cu2(SeO4)2(OH)(H2O)} is isotypic with natrochalcite and consists of CuO$_4$(OH)$_{2}$-based layers interconnected by SeO$_{4}^{2-}$ tetrahedra, with NH$_{4}^{+}$ cations and water molecules occupying the interlayer space~\cite{giester1989crystal}. Natrochalcite-type compounds have recently attracted interest as potential anode materials for Li- and Na-ion batteries~\cite{liu2016evaluation}. \ce{(NH4)2Cu(SeO4)2(H2O)6} is a Tutton salt in which Cu$^{2+}$ is octahedrally coordinated by six water molecules~\cite{monge1981diammonium}, a family of broad current interest for thermochemical energy storage~\cite{smith2024dehydration}, UV-transparent optics in solar-blind detection~\cite{ghosh2018growth}, solid electrolytes~\cite{bejaoui2019synthesis}, fertilizers and waste-recycling by-products~\cite{tastanov2023recycling}, as well as catalytic, antimicrobial, and antioxidant agents~\cite{hfidhi2021lamellar}. In contrast, \ce{(NH4)2Cu(SeO4)2(H2O)2}, although superficially resembling a partially dehydrated Tutton salt, adopts a fundamentally different structure in which \ce{CuO4(H2O)2} octahedra and \ce{SeO4} tetrahedra share corners and alternate along the $a$-axis, rather than being linked solely by hydrogen bonds as in canonical Tutton salts~\cite{fleck2002crystal}.
	
	\begin{figure*}[htb!]
		\begin{center}
			\includegraphics[width=1.0\linewidth]{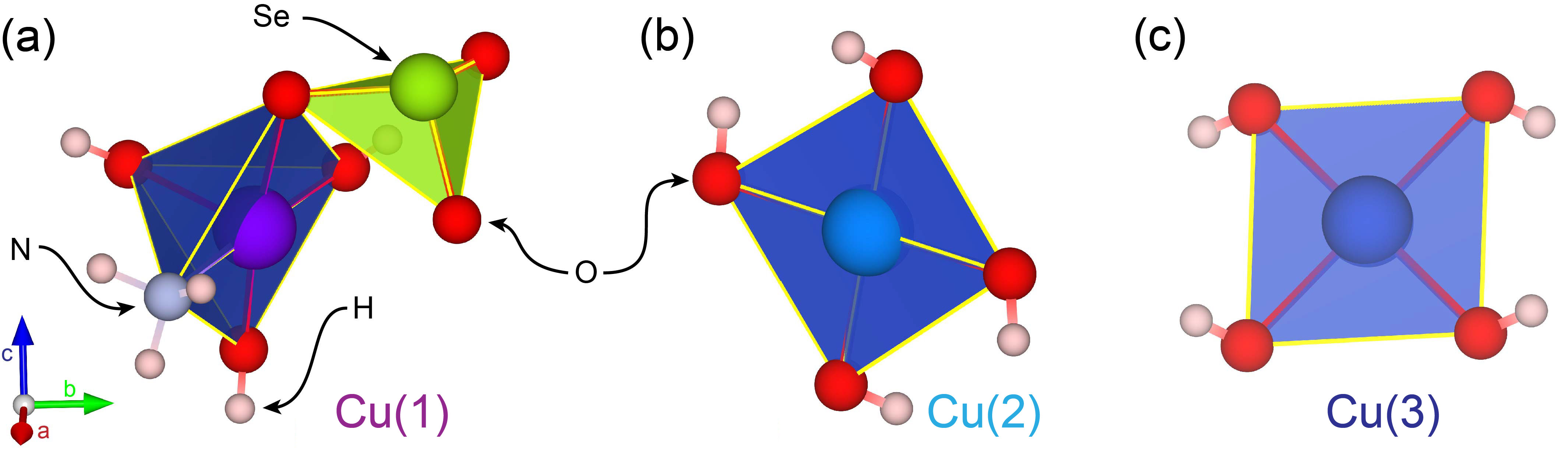}
			\caption{Coordination polyhedra of the (a) Cu(1), (b) Cu(2), (c) Cu(3) atoms in \ce{(NH3)2Cu5(SeO3)2(OH)6\cdot(H2O)_{2+x}} with $x$ = 0, 1, and 3. Copper, selenium, oxygen, nitrogen and hydrogen appear in blue, green, red, light blue and white respectively.}
			\label{fig2}
		\end{center}
	\end{figure*}
	
	Together, these examples show that the Cu–Se–N–O–H system is highly sensitive to hydration state, ammonia versus ammonium incorporation, and selenium oxidation state. They also reveal a striking imbalance: whereas hydrated copper selenates and ammonium-containing phases are represented, hydrated ammine copper selenites remain essentially unexplored. This gap is significant because selenite groups, unlike selenates, introduce stereochemically active lone pairs that can promote low-symmetry architectures, open structural cavities, and unusual hydrogen-bonding environments.
	
	Beyond their structural interest, copper selenites are also attractive because Cu$^{2+}$ carries spin $S$ = 1/2 and can form low-dimensional magnetic lattices. Several copper selenites exhibit unconventional magnetic behavior, including magnetic frustration and quantum spin effects~\cite{constable2017magnetic,badrtdinov2018magnetism,janson2011cacu,seki2012observation,shi2026magnetic}. In such systems, subtle structural modifications can markedly influence magnetic exchange pathways~\cite{arakcheeva2025cu2oseo3,kohn1976crystal,lawes2003magnetodielectric,jeffrey1997introduction}.
	
	\begin{figure}[htb!]
		\begin{center}
			\includegraphics[width=1.00\linewidth]{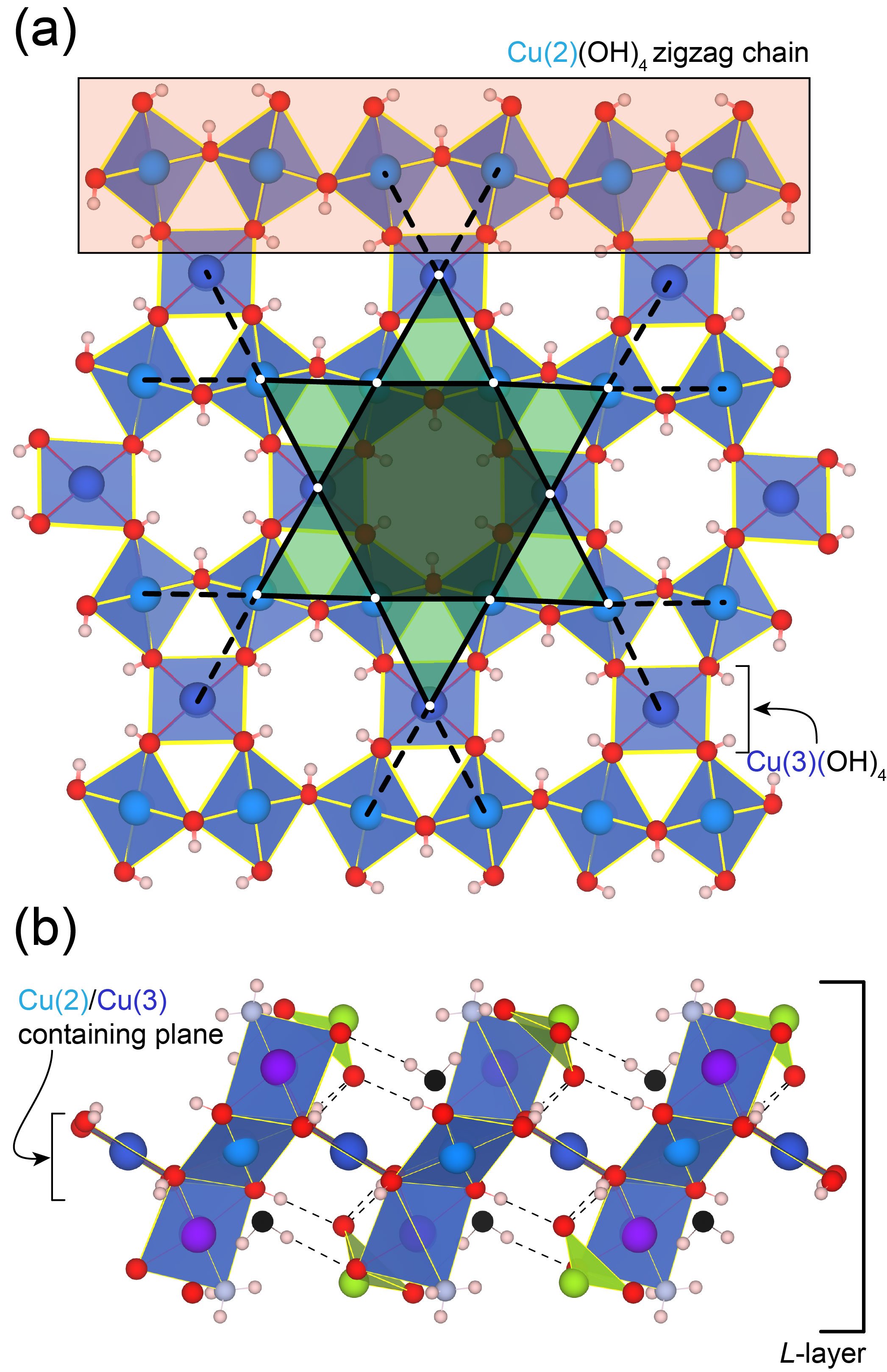}
			\caption{Crystal structure of \ce{(NH3)2Cu5(SeO3)2(OH)6\cdot(H2O)_{2+x}} ($x$ = 1) determined using single crystal X-ray diffraction. (a) Arrangement of the Cu(2) and Cu(3) squares. The zigzag chain of Cu(2)(OH)$_4$ squares is highlighted by an orange rectangle. The zigzag chains are connected by Cu(3)(OH)$_4$ containing squares forming a distorted triangular lattice with both Cu(2) and Cu(3) positioned on the same plane (b) In the $L$-layer, Cu(1) containing square pyramids are present above and below that plane. Intralayer water molecules are shown with oxygen atom coloured in black. (c) The yellow hexagon shows the channels perpendicular to the Cu(2)/Cu(3)-plane (SeO$_3$ groups are omitted for clarity). The presence of Cu1 containing square pyramids are shown as blue and pink triangles for Cu(1) present above and below the Cu(2)/Cu(3) containing planes. Cu(1) containing square pyramids are organized in two triangular lattices (one placed above and one placed below the Cu(2)/Cu(3) containing planes). Cu atoms can be seen as dark blue while red, white, green and light blue atoms are oxygen, hydrogen, selenium and nitrogen, respectively.}
			\label{fig3}
		\end{center}
	\end{figure}
	
	Here, we report the synthesis and structural characterization of a new family of hydrated layered copper selenites, \ce{(NH3)2Cu5(SeO3)2(OH)6\cdot(H2O)_{2+x}} with $x$ = 0, 1, and 3. All three compounds share a common $L$-layer built from Cu(OH)$_4$ squares and Cu-centered square pyramids, forming a distorted kagomé-like Cu$^{2+}$ topology. We show that the degree of hydration controls both the interlayer separation and the relative stacking of adjacent layers, while leaving the intralayer structure essentially unchanged. These compounds therefore provide a rare example of hydration-controlled layer stacking in a transition-metal selenite, in which intercalated water modulates the supramolecular architecture through hydrogen bonding. Finally, we demonstrate that these hydrated layered phases can serve as reactive precursors for the synthesis of the skyrmion-hosting insulator \ce{Cu2OSeO3}.
	
	\vspace{10pt}
	
	\noindent\textbf{Experimental section.} CuSeO$_3\cdot$2H$_2$O is synthesized at room temperature by mixing a concentrated aqueous solution of CuSO$_4\cdot$5H$_2$O with selenious acid prepared by dissolving \ce{SeO2} in water. CuSeO$_3$·2H$_2$O is then filtered and washed with distilled water. For the synthesis of \ce{(NH3)2Cu5(SeO3)2(OH)6\cdot(H2O)_{2+x}} ($x$ = 0, 1, 3), 3.6~g of CuSeO$_3$·2H$_2$O is dispersed in 240~mL of \ce{NH4OH} solution with a concentration from 1.1$\times$10$^{-2}$ to 6.6$\times$10$^{-2}$ molL$^{-1}$. The dispersion is kept between 50~$^\circ$C and 70~$^\circ$C under reflux. Few milliliters of dispersion are regularly collected and quenched at room temperature. The solid is immediately filtered, dried at room temperature without washing such as solids soluble in solvents other than the reaction medium are not lost. Experimental details on powder and single crystal X-Ray diffraction (XRD) as well as three-dimensional electron diffraction (3D-ED) are given in Supplemental Information.
	
	\vspace{10pt}
	
	\noindent\textbf{Results \& Discussion.}\\
	\textit{Synthesis of \ce{(NH3)2Cu5(SeO3)2(OH)6\cdot(H2O)_{2+x}}.} After dispersing CuSeO$_3\cdot$2H$_2$O in the ammonia solution, the round bottom flask is immediately immersed in the silicon oil kept at a temperature between 50~$^\circ$C and 70~$^\circ$C such as the reaction medium reaches readily the growth temperature. A slight color change is observed after few minutes. The blue CuSeO$_3\cdot$2H$_2$O rapidly transforms to turquoise particles (Figure~\ref{fig1}a \& b) while the solution stays transparent which substantiates the absence of dissolved Cu during the reflux treatment. Observation of the sample under optical microscope reveals that the sample is mostly composed of needle like particles with an average size of approximately 20-50 microns (Figure~\ref{fig1}c \& d). The structure of the needles could be determined using single crystal XRD.
	
	\noindent\textit{\ce{(NH3)2Cu5(SeO3)2(OH)6\cdot(H2O)_{2+x}} with $x$ = 1 structure determination.} Sections of reciprocal space are shown in Figure~\textcolor{blue}{S1}. These data confirm a monoclinic lattice with unit-cell parameters $a$ = 22.82(1)~Å, $b$ = 6.50(1)~Å, $c$ = 11.96(2)~Å and $\beta$ = 113.641(11)$^\circ$. The observed systematic absences, $hkl$ with $h + k = 2n$ and $h0l$ with $l = 2n$, are consistent with a $C$-centred monoclinic lattice containing a $c$-glide plane perpendicular to the $b$-axis. The corresponding space group is $C2/c$. The atomic coordinates of the $x = 1$ structure are listed in Table~\textcolor{blue}{S1}, while unit cell parameters are outlined in Table.~\ref{tab:tab1}.
	
	The coordination environments of the Cu atoms are shown in Figure~\ref{fig2}. The Cu(1) atoms adopt a square-pyramidal coordination geometry. In the basal plane of the pyramid, Cu(1) is coordinated by two hydroxo ligands, one ammine ligand, and one selenite oxygen atom. The apical position is occupied by a hydroxo ligand, with a Cu–O distance of approximately 2.20~Å, whereas the two basal Cu–OH distances are shorter, at about 2.00~Å.
	
	The Cu(2)-centered Cu(OH)$_4$ units are non-planar and exhibit $C_1$ point-group symmetry, whereas the Cu(3)-centered Cu(OH)$_4$ units are planar with $C_\mathrm{s}$ symmetry. The Cu–OH distances for both Cu(2) and Cu(3) are slightly shorter than those observed for Cu(1), with average values of 1.956~Å for Cu(2) and 1.938~Å for Cu(3).
	
	The Cu(2)-centered squares share corners and are mutually rotated, forming zigzag chains running along the $b$-axis, as highlighted in orange in Figure~\ref{fig3}a. The Cu(3)-centered squares share corners with the Cu(2)(OH)$_4$, linking these zigzag chains into a pseudo-two-dimensional layer (Figure~\ref{fig3}b). This arrangement generates channels with a hexagonal cross-section, shown in Figure~\ref{fig3}a. These channels are occupied by \ce{SeO3} units (omitted in Figure~\ref{fig3}a), with the Se$^{4+}$ lone pairs oriented toward the center of the channels. The OH groups of the Cu(OH)$_4$ units also point toward the center of the channels and form hydrogen bonds with the \ce{SeO3} groups. The relative arrangement of the Cu(2)- and Cu(3)-centered square units defines a triangular motif reminiscent of a kagomé lattice. This lattice is, however, distorted with Cu(2)– Cu(2)– Cu(3) triangle edge lengths of 3.26, 3.35 and 3.46~Å and internal angles of 57.1$^\circ$, 63.2$^\circ$~ and 59.7$^\circ$.
	
	\begin{figure}[htb!]
		\begin{center}
			\includegraphics[width=1.0\linewidth]{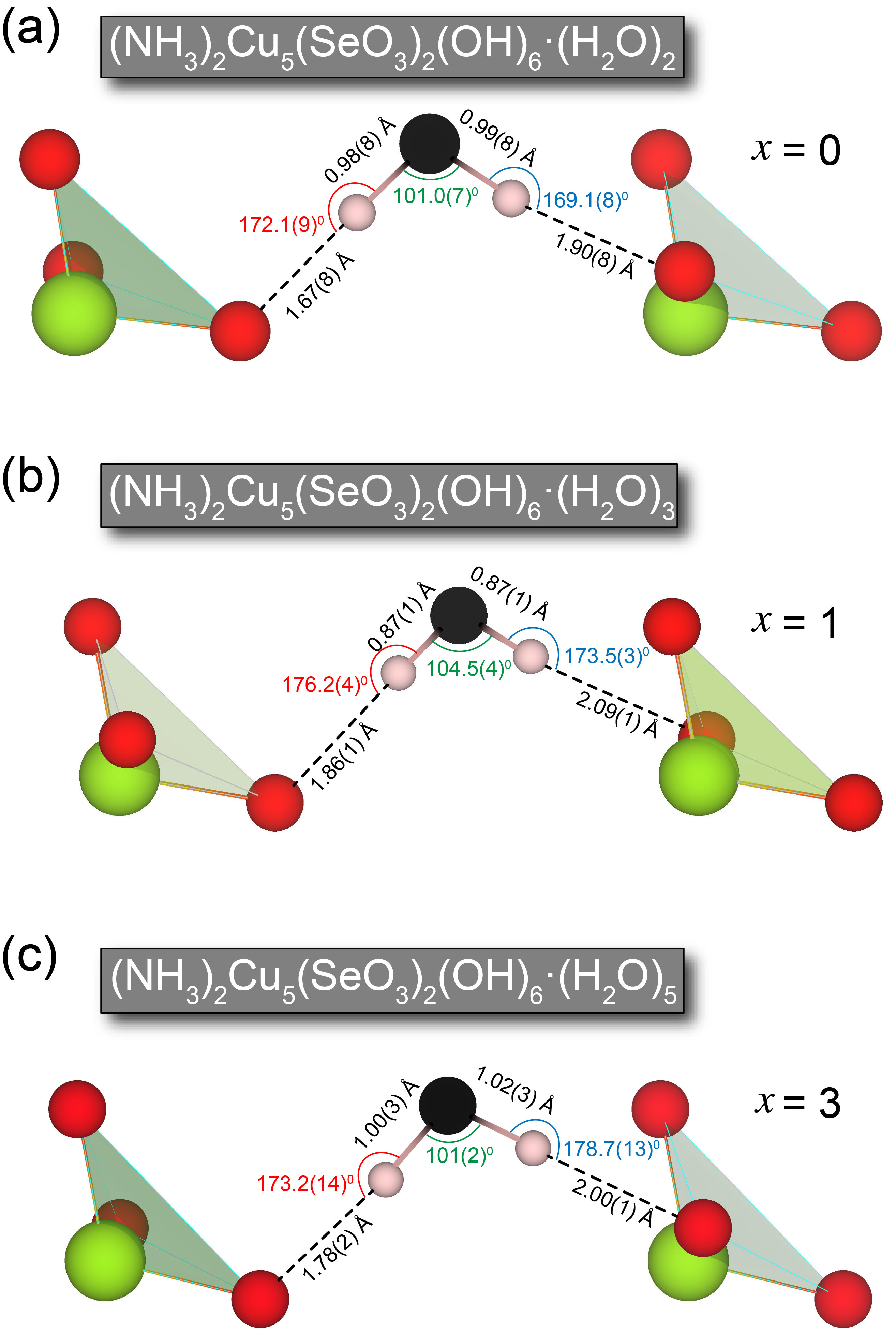}
			\caption{Hydrogen bonding of the intralayer water molecules with the SeO3 groups in (a) $x$ = 0, (b) $x$ = 1 and (c) $x$ = 3. Red, white, and green atoms are oxygen, hydrogen and selenium, respectively. The intralayer oxygen atom is shown in black. See Fig.~\ref{fig3} and Fig.~\ref{fig6} for a complete unit cell visualization. Interatomic distances as well as hydrogen bond angles are given. Symmetry code for this oxygen atom in all three structures is ($x$, $y$, $z$), with a Wycoff position of 4$e$ in $x = 0$ and 8$f$ in $x = 1$ as well as $x = 3$.}
			\label{fig4}
		\end{center}
	\end{figure}
	
	The Cu(1)-centered square pyramids share edges with the Cu(2)- and Cu(3)-centered square-planar units. These pyramids are located above and below the Cu(2)/Cu(3) plane composed of Cu(OH)$_4$ squares, together forming the $L$-layer (Figure~\ref{fig3}a). On each side of the Cu(2)/Cu(3) plane, the Cu(1) pyramids are arranged into two triangular motifs, highlighted in pink and blue, which are rotated by 180$^\circ$ with respect to the Cu(2)/Cu(3) plane. The resulting Cu(1) triangular lattice is distorted, with non-equilateral triangles characterized by edge lengths of 6.50, 6.54 and 7.11~Å and internal angles of 56.7$^\circ$, 57.2$^\circ$ and 66.1$^\circ$.
	
	The water molecules in the structure occupy two distinct crystallographic sites. Two thirds of the water molecules are located within the $L$-layers and are positioned above the centers of the triangles formed by the Cu(2)- and Cu(3)-centered square units, as shown in Figure~\ref{fig3}b and Figure~\ref{fig4}. These water molecules adopt apical positions relative to the Cu(2) and Cu(3) atoms, with Cu$\cdots$O(H$_2$O) distances ranging from 2.5 to 3.0~Å, which are too long for the water molecules to be considered part of the Cu coordination spheres. These \ce{H2O} molecules (H8A–O8–H8B) are instead retained within the $L$-layer through moderate hydrogen bonding interactions with the \ce{SeO3} units (Figure~\ref{fig4}). Although the O–H$\cdots$O hydrogen bonds are nearly linear, with angles of 176–177$^\circ$, the H$\cdots$O distances (1.86–2.08~Å) and O$\cdots$O separations (2.73–2.95~Å) place these interactions in the predominantly electrostatic regime according to Jeffrey’s classification~\cite{jeffrey1997introduction}. The resulting composition of the $L$-layer is therefore \ce{(NH3)2Cu5(SeO3)2(OH)6\cdot(H2O)2}. The remaining one third of the water molecules in the $x = 1$ structure are intercalated between adjacent $L$-layers.
	
	\begin{table*}[htb!]
		\begin{center}
			\caption{Crystallographic details of  \ce{(NH3)2Cu5(SeO3)2(OH)6\cdot(H2O)_{2+x}} ($x$ = 0, 1, and 3). Refinement results for all three structures.}
			\label{tab:tab1}
			\centering
			\begin{tabular}{|c|c|c|c|}
				
				\hline
				
				&$x = 0$&$x = 1$&$x = 3$ \\
				
				& \ce{(NH3)2Cu5(SeO3)2(OH)6\cdot(H2O)2} & \ce{(NH3)2Cu5(SeO3)2(OH)6\cdot(H2O)3} & \ce{(NH3)2Cu5(SeO3)2(OH)6\cdot(H2O)5} \\ \hline
				
				Probe radiation & Electrons ($\lambda$ = 0.02851~Å) & X-ray ($\lambda$ = 1.54184~Å) & Electrons ($\lambda$ = 0.02851~Å) \\ \hline
				
				Crystal size & 0.23 $\times$ 0.38 $\times$ 5.5~$\mu$m$^3$ & 80 $\times$ 40 $\times$ 30~$\mu$m$^3$ & 0.3 $\times$ 1.1 $\times$ 3.3~$\mu$m$^3$ \\ \hline
				
				\textit{T} (K) & 300~K & 139~K & 100~K \\ \hline
				
				$a$, $b$, $c$ (Å) & 10.05(3), 6.53(2), 11.94(4) & 22.82(1), 6.50(1), 11.96(2) & 22.68(3), 6.581(10), 11.945(18) \\ \hline
				
				$\alpha$, $\beta$, $\gamma$ ($^\circ$) & 90, 90.303(8), 90 & 90, 113.641(11), 90 & 90, 93.043(4), 90 \\ \hline
				
				Volume (Å$^3$) & 784(4) & 1624.8(3) & 1780(4) \\ \hline
				
				Space group & $P2_1/c$ (\#14) & $C2/c$ (\#15) & $C2/c$ (\#15) \\ \hline
				
				Formula & \ce{Cu5H16N2O14Se2} & \ce{Cu5H18N2O15Se2} & \ce{Cu5H22N2O17Se2} \\ \hline
				
				$Z$ & 2 & 4 & 4 \\ \hline
				
				Independent reflections & 1264 & 1554 & 1399 \\ \hline
				
				Parameters & 122 & 119 & 122 \\ \hline
				
				Restraints & 6 & 0 & 171 \\ \hline
				
				Resolution (Å) & 0.8 & 0.8 & 0.8 \\ \hline
				
				Completeness (\%) & 78.4 & 99.5 & 76.6 \\ \hline
				
				$R_\text{int}$ (\%) & 12.66 & 3.76 & 7.83 \\ \hline
				
				$R_1$ [$I > 2\sigma(I)$] (\%) & 17.31 & 3.02 & 19.49 \\ \hline
				
				$wR_2$ (all data) (\%) & 44.23 & 7.20 & 46.45 \\ \hline
				
				Goodness of fit & 1.12 & 1.042 & 1.227 \\ \hline
				
				& $-10 \leq h \leq 11$, & $-20 \leq h \leq 28$, & $-26 \leq h \leq 26$, \\
				
				Index ranges & $-8 \leq k \leq 8$, & $-7 \leq k \leq 7$, & $-8 \leq k \leq 8$, \\
				
				& $-14 \leq l \leq 14$ & $-14 \leq l \leq 14$ & $-14 \leq l \leq 14$ \\ \hline
				
				\multirow[c]{2}{8em}{\parbox{8em}{\centering min./max. diff. peaks (e$^-$Å$^{-1}$)}} & & & \\
				& -1.343, 1.273 & 0.794, -0.660 & -1.389, 1.868 \\ \hline
				
			\end{tabular}
		\end{center}
	\end{table*}

\begin{figure}[htb!]
		\begin{center}
			\includegraphics[width=1.0\linewidth]{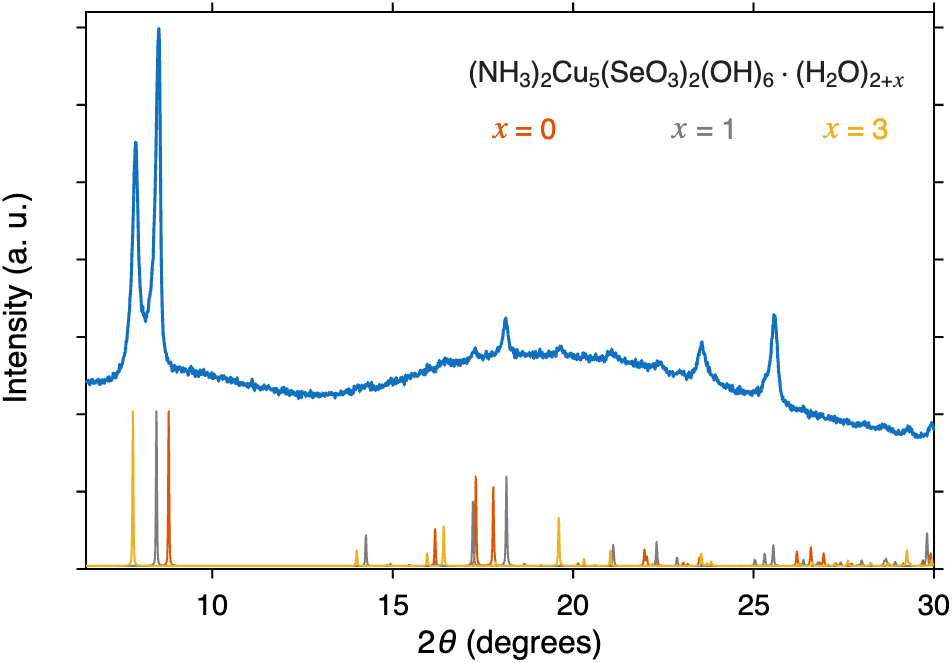}
			\caption{X-ray powder pattern of the \ce{(NH3)2Cu5(SeO3)2(OH)6\cdot(H2O)_{2+x}} ($x$ = 0, 1, and 3). Experimental pattern is shown in blue while the calculated patterns for $x$ = 0, 1 and 3 are shown in grey, orange and yellow respectively.}
			\label{fig5}
		\end{center}
	\end{figure}

	\noindent\textit{\ce{(NH3)2Cu5(SeO3)2(OH)6\cdot(H2O)_{2+x}} with $x = 0$ and 3 structure determination.} As shown in Figure~\ref{fig5}, comparison of the experimental powder X-ray diffraction pattern with the pattern calculated from the crystal structure of the $x = 1$ compound reveals the presence of a reflection at 2$\theta$ = 7.9$^\circ$ that cannot be indexed based on this structure.
	
	3D-ED is a powerful technique for structure determination of crystals that are too small to be analyzed by conventional single-crystal XRD. Moreover, 3D-ED measurements are typically completed within a few minutes, allowing data collection from multiple crystallites. As a result, minor phases present in insufficient quantities to be detected by powder XRD can be identified with ease.
	
	Among the several tens of crystals examined (details provided in Tables~\textcolor{blue}{S2} and \textcolor{blue}{S3}), a small number yielded reciprocal-space reconstructions that did not correspond to the lattice parameters and symmetry of the $x = 1$ structure. Sections of reciprocal space are consistent with a monoclinic lattice with unit-cell parameters $a$ = 10.05(3)~Å, $b$ = 6.53(2)~Å, $c$ = 11.94(4)~Å and $\beta$ = 90.303(8)$^\circ$. The systematic absences are less clear due to dynamical effects, but the reflections conditions $h0l$ with $l = 2n$ and $0k0$ with $k = 2n$ indicate a primitive lattice with a $2_1$ screw axis parallel to the $b$-axis and a $c$-glide plane perpendicular to $b$ (Figure~\textcolor{blue}{S2}). The corresponding space group is $P2_1/c$. The atomic coordinates of the $x = 0$ structure are listed in Table~\textcolor{blue}{S4}.
	
	\begin{figure*}[htb!]
		\begin{center}
			\includegraphics[width=1.0\linewidth]{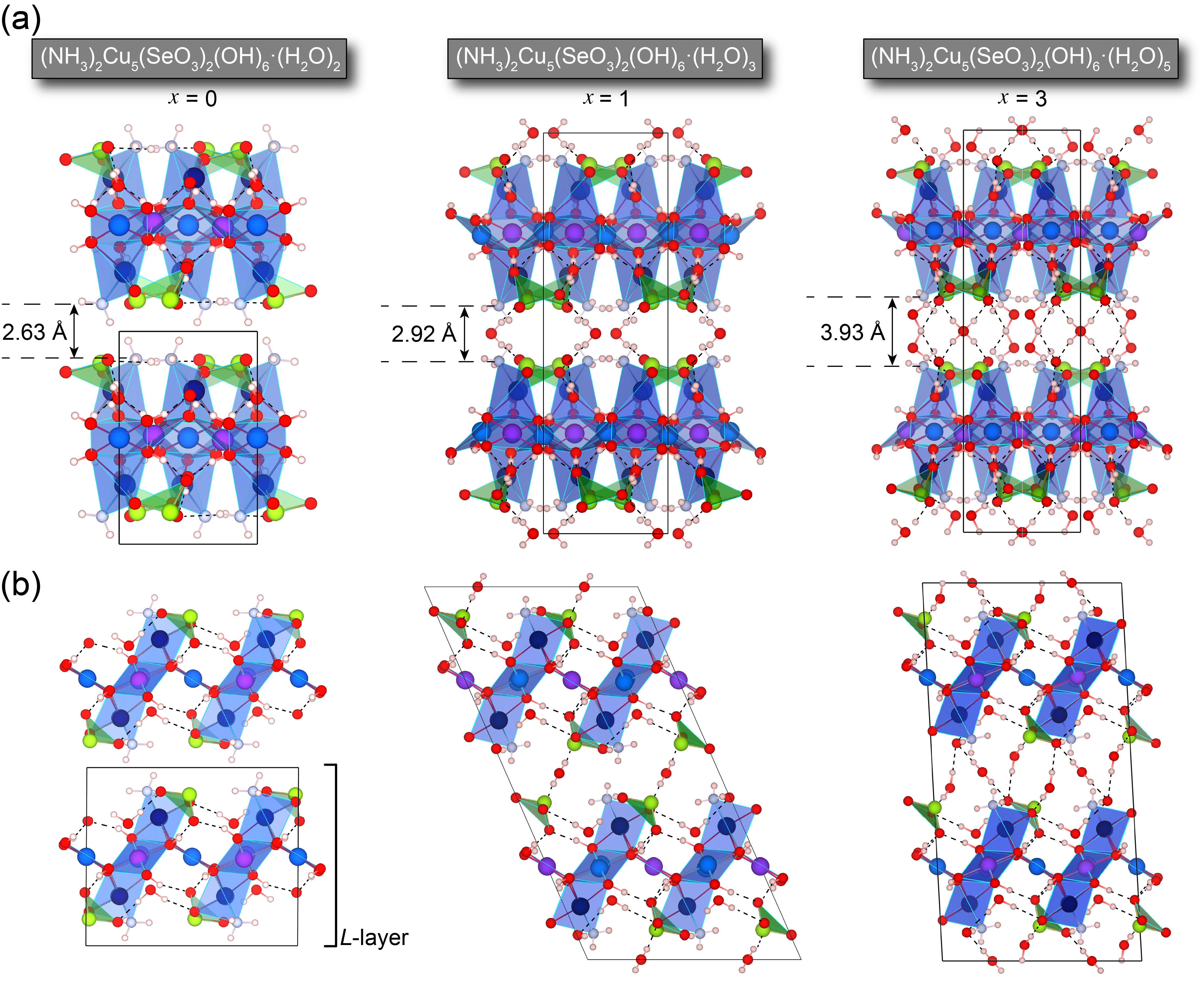}
			\caption{$L$-Layer stacking in \ce{(NH3)2Cu5(SeO3)2(OH)6\cdot(H2O)_{2+x}} with $x$ = 0, 1, and 3. (a) and (b) are orthogonal views of the $L$-Layer. The interlayer distance was measured as the distance between two planes formed by the nitrogen atoms of two successive $L$-layer. Cu atoms can be seen as dark blue while red, white, green and light blue atoms are oxygen, hydrogen, selenium and nitrogen respectively. Note that in \ce{(NH3)2Cu5(SeO3)2(OH)6\cdot(H2O)5}, the proton sites of the water molecules couldn’t be reliably determined. Hence, they were fixed to their most probable position. Single unit cell for each structure is depicted by the solid box. }
			\label{fig6}
		\end{center}
	\end{figure*}
	
	The $x = 0$ structure is built from the same $L$-layers as those observed in the $x = 1$ compound. In the dihydrated ($x = 0$) phase, the Cu(2)/Cu(3) plane is essentially planar, with deviations of approximately 0.05~Å from ideal planarity, comparable to those observed in the trihydrated ($x = 1$) phase. The distorted kagomé lattice formed by the Cu(2) and Cu(3) atoms, as well as the distorted triangular arrangement of the Cu(1) atoms, are identical in the dihydrated and trihydrated structures. In the $x = 0$ structure, the water molecules located within the $L$-layers are stabilized by similar hydrogen-bonding schemes involving neighboring \ce{SeO3} groups, with comparable bond angles, distances and interaction strengths to those observed in the $x = 1$ structure. Consequently, the atomic arrangement within the $L$-layers is not affected by the absence of intercalated water between adjacent $L$-layers.
	
	Among the crystals examined by 3D-ED, many were unstable under the electron beam at room temperature, in contrast to the $x = 0$ and 1 phases. Consequently, further diffraction experiments were carried out at 100~K using a cryogenic sample holder. Sections of reciprocal space exhibit the same systematic absences (albeit less clear due to dynamical effects) as those observed for $x = 1$ structure, indicating an identical space group, although the lattice parameters differ, with $a$ = 22.68(3)~Å, $b$ = 6.581(10)~Å, $c$ = 11.945(18)~Å and $\beta$ = 93.043(4)$^\circ$ (Figure~\textcolor{blue}{S3}). The atomic coordinates of the $x = 3$ structure are listed in Table~\textcolor{blue}{S5}. It should be noted that the hydrogen atoms of the water molecules located within and between the layers could not be reliably determined (further information can be found in Supplementary Materials).
	
	The increased beam sensitivity of the $x = 3$ compound is attributed to the higher water content of the structure. The reflection observed at 2$\theta$ = 7.9$^\circ$ can be indexed based on the $x = 3$ structure. Furthermore, the absence of reflections assigned to $x = 0$ compound indicates that this phase is present only in trace amounts (Figure~\ref{fig5}).
	
	\begin{figure*}[htb!]
		\begin{center}
			\includegraphics[width=1.0\linewidth]{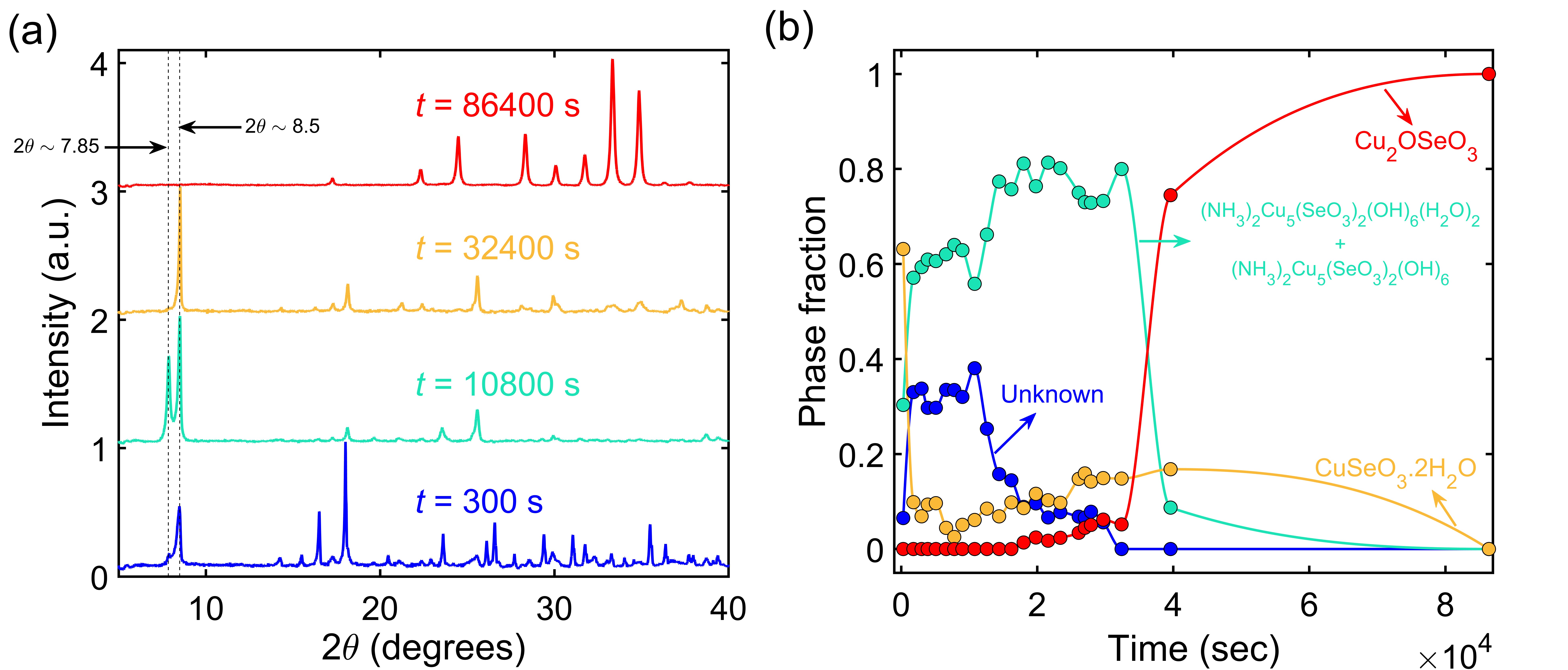}
			\caption{Phase transformation probed using X-ray. (a) Representative powder X-ray patterns. After 5 min, the solid is composed of \ce{(NH3)2Cu5(SeO3)2(OH)6\cdot(H2O)3}, \ce{(NH3)2Cu5(SeO3)2(OH)6\cdot(H2O)5} and CuSeO$_3$.2H$_2$O. After 3~hours, it is composed of \ce{(NH3)2Cu5(SeO3)2(OH)6(H2O)3} and \ce{(NH3)2Cu5(SeO3)2(OH)6(H2O)5}; traces of CuSeO$_3$.2H$_2$O are present. After 9 hours, \ce{(NH3)2Cu5(SeO3)2(OH)6(H2O)3} is the main phase with a small amount of \ce{Cu2OSeO3}. After 24~hours, pure \ce{Cu2OSeO3} is obtained. (b) Phase fraction evolution as a function of reaction time. The phase fraction is calculated as the intensity (peak height) of one specific diffraction peak of the phase (i.e. 35.5$^\circ$, 18$^\circ$, 8.6$^\circ$, and 7.9$^\circ$ in 2$\theta$ for \ce{Cu2OSeO3}, CuSeO$_3$.2H$_2$O, \ce{(NH3)2Cu5(SeO3)2(OH)6(H2O)3} and \ce{(NH3)2Cu5(SeO3)2(OH)6(H2O)5}, respectively) divided by the sum of all the intensities.}
			\label{fig7}
		\end{center}
	\end{figure*}
	
	Like the structure of $x = 0$ and 1, the one of $x = 3$ is built from the same $L$-layers. Although a direct verification of a similar hydrogen-bonding scheme between intralayer water molecules and \ce{SeO3} groups is not possible, the key structural features are preserved. In particular, the distorted kagomé lattice formed by the Cu(2)/Cu(3) atoms and the triangular arrangement of the Cu(1) atoms are identical to those observed in the dihydrated ($x = 0$) and trihydrated ($x = 1$) phases.
	
	\noindent\textit{Layer Stacking in \ce{(NH3)2Cu5(SeO3)2(OH)6\cdot(H2O)_{2+x}} with $x$ = 0, 1, and 3.} In \ce{(NH3)2Cu5(SeO3)2(OH)6\cdot(H2O)_{2+x}}, the incorporation of water molecules between successive $L$-layers has a pronounced effect on the interlayer separation (Figure~\ref{fig6}). In the absence of intercalated water ($x = 0$), the distance between adjacent $L$-layers is 2.63~Å. Upon insertion of one additional water molecule ($x = 1$), the separation increases to 2.92~Å. Further hydration ($x = 3$) leads to a substantial expansion of the interlayer spacing to 3.93~Å, corresponding to an increase of approximately 50\% relative to the anhydrous interlayer configuration ($x = 0$).
	
	Beyond the pronounced expansion of the interlayer distance, the presence or absence of water also induces significant changes in the stacking arrangement. The relative arrangement of the $L$-layers differs depending on the presence of intercalated water. In the absence of intercalated water ($x = 0$), successive $L$-layers are shifted by approximately half a unit cell (Figure~\ref{fig6}a), such that the \ce{NH3} groups of one layer face the \ce{SeO3} groups of the adjacent $L$-layer. In contrast, when water is present ($x$ = 1 and 3), the stacking along the $c$-axis results in \ce{NH3} groups facing \ce{NH3} groups in neighboring $L$-layers. Along the second in-plane direction, the relative positioning of the $L$-layers remains unchanged (Figure~\ref{fig6}b). For $x = 1$, the interlayer water molecules are located on a single plane situated approximately at the center of the interlayer space, whereas for $x = 3$ they are distributed over three approximately equidistant planes. In the $x = 1$ structure, the water molecules are positioned between two facing \ce{SeO3} groups and form hydrogen bonds with geometrical parameters (O–H$\cdots$O angle = 175$^\circ$, H$\cdots$O distance = 2.06~Å and O···O distance = 2.91~Å) comparable to those observed for water molecules within the $L$-layers. These intercalated water molecules therefore contribute to the cohesion between successive $L$-layers. For $x = 3$, the hydrogen atoms of the intercalated water molecules could not be located, and thus the presence and nature of hydrogen-bonding interactions involving the three distinct interlayer water sites cannot be directly established. However, the O$\cdots$O distance between the water molecules located in the middle plane of the interlayer region and the nearest \ce{SeO3} oxygen atoms (2.67~Å) suggests the possibility of hydrogen bonding between adjacent $L$-layers, contributing to its structural stability. The two other interlayer water sites are located closer to one $L$-layer and farther from the next (O$\cdots$O distances of 2.78 and 3.04~Å, respectively), indicating a weaker contribution to interlayer cohesion compared with the centrally located water molecules.
	
	\noindent\textit{Conversion of \ce{(NH3)2Cu5(SeO3)2(OH)6\cdot(H2O)_{2+x}} with $x$ = 0, 1, 3 into \ce{Cu2OSeO3}.} As shown in Fig.~\ref{fig1}, \ce{(NH3)2Cu5(SeO3)2(OH)6\cdot(H2O)_{2+x}} phases are obtained by treating CuSeO$_3$·2H$_2$O in an ammoniacal solution, according to the reaction:
	
	The leached selenious acid is present as SeO$_{3}^{2-}$ in the reaction medium, as the pK$_\text{a}$ of ammonia (9.23) exceeds those of diprotic selenious acid (pK$_\text{a1}$ = 2.62 and pK$_\text{a2}$ = 8.32). When \ce{(NH3)2Cu5(SeO3)2(OH)6\cdot(H2O)_{2+x}} is maintained under reflux conditions, and the initial stoichiometry between CuSeO$_3$·2H$_2$O and \ce{NH4OH} is 1:1, it is progressively converted into \ce{Cu2OSeO3}. The evolution of the solid-phase composition during reflux was monitored by powder XRD (Figure~\ref{fig7}a). Owing to the needle-like morphology of \ce{(NH3)2Cu5(SeO3)2(OH)6\cdot(H2O)_{2+x}}, which induces strong preferred orientation, the phase fractions should be regarded as qualitative.
	
	After 5~min of reaction, CuSeO$_3$·2H$_2$O is converted into $x = 1$ and 3 materials, with traces of the $x = 0$ phase detected only by electron diffraction. The fractions of the $x = 1$ and $x = 3$ phases increase progressively until reaching a plateau after approximately 2~hours (Figure~\ref{fig7}b). The release of SeO$_{3}^{2-}$ from CuSeO$_3$·2H$_2$O modifies the composition of the reaction medium. Given that the $x = 1$ and 3 compounds exhibit a higher Cu/Se ratio (2.5) than \ce{Cu2OSeO3} (Cu/Se = 2), the formation of \ce{Cu2OSeO3} proceeds via reaction of these intermediate phases with the selenite species released into the solution:
	
	After 24~h of reflux, both precursor and intermediate phases are fully converted into \ce{Cu2OSeO3}. The powder XRD pattern confirms the formation of a phase-pure \ce{Cu2OSeO3} sample (Figure~\ref{fig7}a). Bulk magnetometry as well as small-angle neutron scattering results on thus synthesized \ce{Cu2OSeO3} particles are reported elsewhere~\cite{baral2022tuning}.
	
	\vspace{10pt}
	
	\noindent\textbf{Conclusion.} A new structural family of layered copper selenites, \ce{(NH3)2Cu5(SeO3)2(OH)6\cdot(H2O)_{2+x}} ($x$ = 0, 1, 3), has been successfully synthesized and characterized. All members share a common $L$-layer framework based on Cu$^{2+}$ polyhedra forming distorted kagomé-like arrangements. While the intralayer structure remains essentially unchanged across the series, the degree of hydration has a profound impact on both the interlayer distance and relative orientation of adjacent layers, identifying these compounds as new candidate platforms for hydration-responsive materials. Furthermore, these compounds have been shown to serve as intermediates in the formation of \ce{Cu2OSeO3}, providing a new synthetic pathway toward this skyrmion-host material. 
	
	From a broader perspective, \ce{(NH3)2Cu5(SeO3)2(OH)6\cdot(H2O)_{2+x}} may represent the first member of a wider structural family of the type \ce{(NH3)2\textit{M}5(SeO3)2(OH)6\cdot(H2O)_{2+x}} ($M$ = Zn, Mg, Mn, Fe, Co, Ni), potentially accessible from known hydrated or anhydrous transition-metal selenites. Substituting Cu$^{2+}$ ($S$ = 1/2) with other divalent transition-metal ions such as Ni$^{2+}$ ($S$ = 1), Co$^{2+}$ ($S$ = 3/2), Mn$^{2+}$ ($S$ = 5/2), or Fe$^{2+}$ ($S$ = 5/2) would offer promising opportunities to explore magnetic interactions in a potentially frustrated lattice. Such extensions could provide a versatile platform for investigating the interplay between structure, hydration, and magnetism in low-dimensional systems.
	
	\vspace{10pt}
	
	\noindent \textbf{Author contribution} AM supervised the project. PRB, PP, JR and AM synthesized the samples, WHB performed the single crystal XRD and refined the structure. CJ performed the 3D-ED experiments and refined the structures. AM and PRB analyzed the data. AM and PRB wrote the first draft of the manuscript. All authors read, commented and revised the manuscript.
	
	\vspace{10pt}
	
	\noindent \textbf{Acknowledgments} This work was supported by the Swiss National Science Fundation (SNSF) Sinergia network NanoSkyrmionics (grant No. CRSII5-171003). Additionally, P.R.B. acknowledges SNSF Return CH Postdoc.Mobility grant P5R5-2\_239356 for financial support.
	
	\vspace{10pt}
	
	\noindent\textbf{References}
	
	\bibliography{biblio}

\end{document}